\documentclass[showpacs,preprintnumbers,amsmath,amssymb,twocolumn]{revtex4-1}
\usepackage{graphicx}
\usepackage{dcolumn}
\usepackage{bm}
\usepackage{subfigure}
\usepackage{float}
\usepackage{multirow}
\usepackage{booktabs}
\usepackage{amsmath}
\usepackage{latexsym,amssymb}
\usepackage[mathscr]{eucal}
\usepackage[switch*]{lineno}
\usepackage[bookmarks=true,
   colorlinks=true,
   linkcolor=blue,
   urlcolor=blue,
   citecolor=blue,
   bookmarks=true,
   hyperindex=true
]{hyperref}

\begin{document}

\title{Higher-order generalized uncertainty principle corrections to the Jeans mass}

\author{Zhong-Wen Feng\textsuperscript{1}}
\altaffiliation{Email: zwfengphy@cwnu.edu.cn}
\author{Guansheng He\textsuperscript{2}}
\author{ Xia Zhou\textsuperscript{1}}
\author{Xueling Mu\textsuperscript{3}}
\author{Shi-Qi Zhou\textsuperscript{1}}
\vskip 1cm
\affiliation{1. Physics and Space Science College, China West Normal University, Nanchong, 637009, China\\
2. School of Mathematics and Physics, University of South China, Hengyang 421001, China\\
3. School of Information Science and Engineering, Chengdu University, Chengdu 610106, China}

\date{\today}

\begin{abstract}
The Jeans instability is regarded as an important tool for analyzing the dynamics of a self-gravitating system. However, this theory is challenging since astronomical observation data show some Bok globules, whose masses are less than the  Jeans mass and still have stars or at least undergo the star formation process. To explain this problem, we investigate the ef\/fects of the higher-order generalized uncertainty principle on the Jeans mass of the collapsing molecular cloud. The results in this paper show that the higher order generalized uncertainty principle has a very signif\/icant ef\/fect on the canonical energy and gravitational potential of idea gas, and f\/inally leads to a modif\/ied Jeans mass lower than the original case, which is conducive to the generation of stars in small mass Bok globules. Furthermore,  we estimate the new generalized uncertainty principle parameter $\gamma _0$ by applying various data of Bok globules, and f\/ind that the range of magnitude of  $\gamma _0$ is ${10^{11}} \sim {10^{12}}$.
\end{abstract}
\maketitle
\section{Introduction}
\label{Int}
Einstein's theory of general relativity is regarded as the cornerstone of the development of modern physics and astronomy. However, with the deepening of research, it is found that general relativity has f\/laws, which lead to many problems, such as the black hole information paradox and naked singularity of spacetimes \cite{chb1}.  One of the most promising candidates to solve those problems is quantum gravity (QG). Now, based on these frameworks that explicate  QG, people hypothesize that exists a minimum measurable length of the order of the Planck length in QG, which is supported by Gedanken experiments \cite{cha1,cha2,cha3}.

According to the minimum measurable length, many such studies have converged on the idea that the Heisenberg uncertainty principle (HUP) can be modif\/ied as the generalized uncertainty principle (GUP). In 1995, Kempf, Mangano and Mann proposed a quadratic form of the GUP,  which we now call as the ``KMM model" with the  expression, $\Delta x\Delta p \ge \frac{\hbar }{2}\left[ {1 + \beta  \Delta {p^2}} \right]$, where $\beta  = {{{\beta  _0}} \mathord{\left/ {\vphantom {{{\beta  _0}} {M_p^2{c^2}}}} \right. \kern-\nulldelimiterspace} {M_p^2{c^2}}} = {{{\beta_0}\ell _p^2{\rm{ }}} \mathord{\left/  {\vphantom {{{\beta_0}\ell _p^2{\rm{ }}} {{\hbar ^2}}}} \right. \kern-\nulldelimiterspace} {{\hbar ^2}}}$ and $\beta  _0$ is the GUP parameter. It is easy to f\/ind that the KMM model predicts a minimal length $\Delta x_{\min }^{{\rm{KMM}}} \approx {\ell _p}\sqrt {{\beta_0}}$ \cite{cha4}. Subsequently, incorporating the idea of maximal momentum, Ali, Das and Vagenas constructed another GUP (ADV model)  $\Delta x\Delta p \ge \frac{\hbar }{2}\left[ {1 + {{2{\alpha _0}{\ell _p}\left\langle p \right\rangle } \mathord{\left/
 {\vphantom {{2{\alpha _0}{\ell _p}\left\langle p \right\rangle } \hbar }} \right. \kern-\nulldelimiterspace} \hbar }} + {  {{4\left\langle {{p^2}} \right\rangle \alpha _0^2\ell _p^2} \mathord{\left/ {\vphantom {{4\left\langle {{p^2}} \right\rangle \alpha _0^2\ell _p^2} {{\hbar ^2}}}} \right. \kern-\nulldelimiterspace} {{\hbar ^2}}}} \right]$, where $\alpha_0$ is the GUP parameter. Moreover, this linear and quadratic GUP model suggests the existence of a minimal length $\Delta x_{\min }^{{\rm{ADV}}} \approx {\alpha _0}{\ell _p}$ and a maximal momentum  $\Delta p_{\max }^{{\rm{ADV}}} \approx {{{\ell _p}} \mathord{\left/ {\vphantom {{{\ell _p}} {{\alpha _0}}}} \right. \kern-\nulldelimiterspace} {{\alpha _0}}}$\cite{cha5}. In recent years, the KMM model and ADV model have played important roles in the research of many physics contexts. For instance, in Refs.~\cite{cha5+,cha5++}, the Scardigli \emph{et al.}, discussed the relation between GUP, general relativity, and the Lorentz violation. Besides, by using the KMM model, researchers derived the GUP corrected Hamilton-Jacobi equation and  investigated the modif\/ied tunneling rate of particles with arbitrary spins from the event of curved spacetimes \cite{cha6}. Moreover, the KMM model and ADV model can also be extended to the Proca equation, which leads to the modif\/ied Hawking temperature of black holes \cite{cha7,cha8,cha8+}. In addition, Vagenas \emph{et al}. studied the validity of the no-cloning theorem within the framework of GUP. They pointed out that the energy required to send information to a black hole is af\/fected by quantum gravity \cite{cha9}. In Refs.~\cite{cha10,cha10+}, the authors investigated how the KMM model af\/fects the Casimir wormhole spacetime, and obtained a class of asymptotically f\/lat wormhole solutions.

Although the KMM model and ADV model are two of the most studied GUPs, they still have some defects, e. g., the minimal length and the maximal momentum are only valid for small GUP parameters, and do not imply noncommutative geometry \cite{cha11}. To overcome these dif\/f\/iculties, Pedram introduced a nonperturbative higher-order GUP, which agrees with various proposals of QG \cite{cha12}. Subsequently, this higher order GUP was used to correct the blackbody radiation spectrum and predict the cosmological constant. Based on this heuristic work, many new forms of higher order GUP were presented (see, e. g.  Refs.~\cite{cha13,cha14,cha15,cha15+} and references therein). Recently, in Ref.~\cite{cha16},  Chung and Hassanabadi proposed a generalized canonical commutation relations as follows:
\begin{align}
\label{eq1}
[\hat x,\hat p] = \frac{{i\hbar }}{{1 - \gamma \left| {\hat p} \right|}}, \quad \gamma  > 0
\end{align}
which leads to a new higher order GUP (hereafter, we call this GUP as the CH model):
\begin{align}
\label{eq2}
\Delta x\Delta p & \ge  \frac{\hbar }{2}\left\langle {\frac{1}{{1 - \gamma \left| p \right|}}} \right\rangle
\nonumber \\
& =  \frac{\hbar }{2}\left[ { - \gamma \left( {\Delta p} \right) + \frac{1}{{1 - \gamma \left( {\Delta p} \right)}}} \right],
\end{align}
where $\left| p \right| = \sqrt {\left| {{p^2}} \right|} $ and $\gamma  = {{{\gamma _0}} \mathord{\left/ {\vphantom {{{\gamma _0}} {{M_p}c}}} \right. \kern-\nulldelimiterspace} {{M_p}c}}= {{{\gamma  _0}\ell _p{\rm{ }}} \mathord{\left/ {\vphantom {{{\gamma  _0}\ell _p{\rm{ }}} {{\hbar }}}} \right. \kern-\nulldelimiterspace} {{\hbar}}}$ are associated with the dimensionless GUP parameter $\gamma _0$, the Planck mass  $M_p$, and the Planck length  ${\ell _p}$. In addition, we set $\left\langle p \right\rangle  = 0$, $\left\langle {\left| p \right|} \right\rangle  \ge 0$ and utilized the identities $\left\langle {{{\left( {{p^2}} \right)}^n}} \right\rangle  \ge {\left( {\left\langle {{p^2}} \right\rangle } \right)^n}$ and $\left| {\left\langle {\left( {AB + BA} \right)} \right\rangle } \right| \ge 2\sqrt {\left\langle {{A^2}} \right\rangle } \sqrt {\left\langle {{B^2}} \right\rangle }$  \cite{cha16}. In addition, Eq.~(\ref{eq2}) guarantees the existence of an absolute smallest uncertainty in position $\Delta {x_{\min }} = {{3\hbar \gamma } \mathord{\left/ {\vphantom {{3\hbar \gamma } 2}} \right. \kern-\nulldelimiterspace} 2}$  for $\Delta p = {1 \mathord{\left/  {\vphantom {1 {2 \gamma}}} \right.  \kern-\nulldelimiterspace} {2 \gamma }}$, which never appears in the framework of the HUP. In addition, the CH model has merit since it involves no perturbation and overcomes some conceptual problems raised in the previous forms of GUP (e.g. the divergence of the energy spectrum of the eigenfunctions of the position operator), however, when $\gamma  \rightarrow 0$, it reduces to the HUP. Besides, for $\gamma\ll1$, the last term of Eq.~(\ref{eq2}) can be expanded as $\Delta x\Delta p \ge \frac{\hbar }{2}\left[ {1 + {\gamma ^2}\Delta {p^2} + \mathcal{O}\left( {{\gamma ^3}} \right)} \right]$, which indicates that the CH model coincides with the KMM model to the second order in $\gamma$; hence, one has the relation ${\gamma ^2} \sim \beta$.

 According to the CH model~(\ref{eq2}), people investigated the modif\/ied eigenfunctions and eigenvalues for the particle in a box and one-dimensional hydrogen atom, respectively. Moreover, one may also f\/ind that the higher order GUP corrected the behavior of Bohr-Sommerfeld quantization, which can be used to calculate the energy spectra of quantum harmonic oscillators and quantum bouncers \cite{cha15}. In light of previous works, it is believed that a higher-order GUP will not only give new implications to quantum systems, but also to astrophysics. For example, it is well known that the Jeans mass is an important theory to study the collapse of molecular clouds. However, in recent years, this theory is has been since some astronomical data have shown that there are some Bok globules, whose masses are less than the  Jeans mass that still have stars or at least undergo the star formation process \cite{cha17,cha18}.  For solving this problem, Moradpour  \emph{et al}.  corrected the limit of Jeans mass due to the KMM model. Their results showed that the modif\/ications were less than the original case and were related to the properties of the GUP model \cite{cha19}. As we discussed above, the higher order GUP has  dif\/ferent properties from the KMM model and the ADV model, and it is interesting to investigate how the higher order GUP af\/fects the Jeans mass. Therefore,  in the present paper, we try to extend the CH model~(\ref{eq2}) into the Jeans instability and calculate the GUP corrected limit of Jeans mass of those especial Bok globules.

 The paper is organized as follows: Section~\ref{sec2} is devoted to a review of the Jeans instability limit in the classical case. In Section~\ref{sec3}, according to the new higher order CH model, we compute the corrections to the potential energy and canonical energy of a molecular cloud. Then, we discuss the modif\/ied Jeans gravitational instability, and then derive the GUP corrected Jeans mass to explain the problem of Bok globules. By using the GUP corrected Jeans mass, the dimensionless parameter $\gamma_0$ of the GUP is constrained in Section~\ref{sec4}. Finally, conclusions can be found in Section~\ref{sec5}.

\section{The Jeans mass and the HUP}
\label{sec2}
In this section, we brief\/ly  outline how to  derive the Jeans mass in the classical case. Based on the argument that relies upon the virial theorem in Ref.~\cite{cha19}, the collapse of the molecular cloud occurs if the gravitational potential energy $E_p$ is larger than the canonical energy $U$, to wit
 \begin{align}
\label{eq3}
 U <  - {E_p \mathord{\left/ {\vphantom {E_p 2}} \right. \kern-\nulldelimiterspace} 2}.
\end{align}
Obviously, to obtain the Jeans mass, it is necessary to derive the expressions of gravitational potential energy and canonical energy.

It is well known that spacetime is flat in the classical case. Therefore, the Newton's law of gravitation is  ${F_0} = {{GMm} \mathord{\left/ {\vphantom {{GMm} {{r^2}}}} \right. \kern-\nulldelimiterspace} {{r^2}}}$ with the constant of universal gravitation $G$, and the gravitational potential becomes  $V\left( r \right) =  - {{GM} \mathord{\left/ {\vphantom {{GM} r}} \right. \kern-\nulldelimiterspace} r}$. The corresponding potential energy is given by
 \begin{align}
\label{eq4}
E_p & = \int_0^M {V\left( r \right)dr}  = \int_0^R {4\pi \rho \left( r \right)rGMdr}
\nonumber \\
& =  - \frac{{3G{M^2}}}{{5R}},
\end{align}
where the density of dark cloud $\rho \left( r \right)$  in Eq.~(\ref{eq4}) is assumed to be a constant  ${\rho _0}$ for simplicity.

Next, the classical fundamental commutation relation can be expressed as $[\hat x,\hat p] = i\hbar$, which implies HUP $\Delta x\Delta p \ge {\hbar  \mathord{\left/  {\vphantom {\hbar  2}} \right. \kern-\nulldelimiterspace} 2}$. When considering the Liouville theorem and HUP, the state density in momentum space in the spherical coordinate system can be expressed as $D\left( p \right)dp = {{4\pi V{p^2}dp} \mathord{\left/ {\vphantom {{4\pi V{p^2}dp} {{h ^3}}}} \right. \kern-\nulldelimiterspace} {{h^3}}}$, which leads directly to the original partition function of  the ideal gas as follows
\begin{align}
\label{eq5}
Z &= \frac{{\mathcal{Z}^N}}{{N!}} = \frac{{{{\left( {4\pi V} \right)}^N}}}{{N!}}\int {{{\left[ {\frac{{{p^2}}}{{{h^3}}}\exp \left( { - \frac{{{p^2}}}{{2\mu {k_B}T}}} \right)} \right]}^N}dp}
\nonumber \\
 &=\frac{{{{\left( {4\pi V} \right)}^N}}}{{N!}}\frac{{{{\left( {2\pi \mu {k_B}T} \right)}^{3N/2}}}}{{{h^{3N}}}},
\end{align}
where  $N$, $\mu$, and $T$  are the numbers, mass, and temperature of noninteracting particles of  the ideal gas, respectively \cite{che1,cha20,cha21}. Besides, the total mass of the ideal gas satisf\/ies the relationship $M=\mu N$. According to Eq.~(\ref{eq5}), one can straightforward to derive the equation of state of the ideal gas $P V=N k_B T$, and the canonical energy can be expressed as
\begin{align}
\label{eq7}
U_0  = {k_B}{T^2}{\left( {\frac{{\partial \ln Z}}{{\partial T}}} \right)_{N,V}} = \frac{3}{2}N{k_B}T.
\end{align}
Clearly, the original canonical energy is dependent only on the number of noninteracting particles $N$  and their temperature  $T$.

Now, substituting Eq.~(\ref{eq4}) and Eq.~(\ref{eq7}) into inequality~(\ref{eq3}), and supposing that the distribution of ideal gas in space is spherical, the result is
\begin{align}
\label{eq8}
M > {\left( {\frac{{5{k_B}T}}{{G\mu }}} \right)^{\frac{3}{2}}}{\left( {\frac{3}{{4\pi {\rho _0}}}} \right)^{\frac{1}{2}}},
\end{align}
where the radius is $R = {\left( {{{3M} \mathord{\left/ {\vphantom {{3M} {4\pi {\rho _0}}}} \right. \kern-\nulldelimiterspace} {4\pi {\rho _0}}}} \right)^{\frac{1}{3}}}$.  The above mentioned equation indicates a lower bound of the cloud mass to collapse, which is known as the Jeans mass
\begin{align}
\label{eq8+}
M_0^J = {\left( {\frac{{5{k_B}T}}{{G \mu}}} \right)^{\frac{3}{2}}}{\left( {\frac{3}{{4\pi {\rho _0}}}} \right)^{\frac{1}{2}}}.
\end{align}
Although the formation of most stars can be examined by the original Jeans instability and Jeans mass, people still observed some Bok globules, such as CB 84 and CB 110, have masses less than ${M^J_0}$ but still have stars or at least undergo the star formation process  \cite{cha18}. To explain this problem, we investigate the modif\/ied Jeans mass in the framework of the higher-order GUP.

\section{The Jeans mass in the higher-order GUP}
\label{sec3}
The GUP has various implications for a wide range of physical systems. Moreover, they are regarded as a powerful tool to solve various dif\/f\/icult problems in these research f\/ields. Therefore, to obtain the modif\/ied Jeans mass, we derive the GUP corrected entropy,  gravitation potential energy, and canonical energy of ideal gases. For the sake of simplicity, we shall takes the units $\hbar=c=k_B=1$ in the next research.

\subsection{GUP corrected gravitation potential energy}
To derive the modif\/ied gravitation potential energy in the presence of the CH model, we need to use the theory of entropic force, which  is an intriguing explanation for Newton's law of gravity that is based on the holographic principle and an equipartition rule. Now, by using Eq.~(\ref{eq2}), one can easily obtain the following inequality
\begin{align}
\label{eq9}
\Delta p  & \ge \frac{1}{{2\gamma }}\left( {1 - \sqrt {1 - \frac{{4\gamma }}{{2\Delta x + \gamma }}} } \right)
\nonumber \\
& = \frac{1}{{2\Delta x}}\left[ {1 + \frac{{{\gamma ^2}}}{{4\Delta {x^2}}} + \mathcal{O}\left( {{\gamma ^3}} \right)} \right].
\end{align}
Based on the arguments of Refs.~\cite{cha23,chd1,chd2,chd3,chd3+}, a massless quantum particle  (e.g., a photon) can be used to determine the position of quantum particles with the energy $\omega  = pc$ (i.e. $\Delta \omega  = \Delta p c$), so that the HUP $\Delta p \ge {1 \mathord{\left/ {\vphantom {1 {2\Delta x}}} \right. \kern-\nulldelimiterspace} {2\Delta x}}$  can be translated to the lower bound $\omega  \ge {1 \mathord{\left/ {\vphantom {1 {2\Delta x}}} \right.
 \kern-\nulldelimiterspace} {2\Delta x}}$. Accordingly, from Eq.~(\ref{eq9}),  the GUP version reads
\begin{align}
\label{eq11}
\omega  \ge \frac{1}{{2\Delta x}}\left[ {1 + \frac{{{\gamma ^2}}}{{4\Delta {x^2}}} + \mathcal{O}\left( {{\gamma ^3}} \right)} \right].
\end{align}
From Verlinde's entropy force theory, to calculate the modif\/ied entropic force, one should consider a spherically symmetric gravitational system (e.g., a black hole), which allows the quantum particles to enter or exit its horizon \cite{cha28}. As Ref.~\cite{chy1} pointed out, when a strong gravitational system absorbs  or emits a quantum particle with energy $\omega$ and size $\mathcal{R}$, the minimal change in the area of the system becomes $\Delta {A_{\min }} \ge 8\pi  \omega \mathcal{R} \ell _p^2$. Then, the arguments of Refs.~\cite{cha24,chy2} implies that the size of a quantum particle cannot be smaller than its uncertainty in the position (i.e. $\mathcal{R} \geq \Delta x$),  which gives $\Delta {A_{\min }} \ge 8\pi \omega \Delta x \ell _p^2$. Substituting this relation into Eq.~(\ref{eq11}), one has
\begin{align}
\label{eq11+}
\Delta {A_{\min }} \ge 4\pi \ell _p^2\left[ {1 + \frac{{{\gamma ^2}}}{{4\Delta {x^2}}} + \mathcal{O}\left( {{\gamma ^3}} \right)} \right].
\end{align}
Furthermore, considering that a spherical gravitational system with Schwarzschild radius $r$, and the area of this system is $A = 4\pi {r^2}$. Moreover, near the event horizon, $\Delta x$ is approximately equal to the orbital radius $r$; that is, $\Delta x \approx 2r$ \cite{cha25,cha26,chy3,chy4,chy4+}. Hence, the relation between $\Delta x$ and $A$ is given by  $\Delta {x^2} = 4 {r^2} = {A \mathord{\left/ {\vphantom {A \pi }} \right. \kern-\nulldelimiterspace} \pi }$. Substituting this relation into Eq.~(\ref{eq11+}), the minimal change in the horizon area of the gravitational system can be recast as
\begin{align}
\label{eq12}
\Delta {A_{\min }} \simeq \lambda \ell _p^2\left[ {1 + \frac{{\pi {\gamma ^2}}}{{4A}} + \mathcal{O}\left( {{\gamma ^3}} \right)} \right],
\end{align}
where  $\lambda$ is an undetermined coef\/f\/icient. In Ref.~\cite{cha24}, the authors pointed out that the information of the gravitational system is ref\/lected in its area. On the other hand, based on the information theory, it is believed that the area of one system can af\/fect its smallest increase in entropy \cite{cha27}. Since the fundamental unit of entropy as one bit of information is $\Delta {S_{\min }} = b = \ln 2$, one obtains
\begin{align}
\label{eq12+}
\frac{{{\rm{d}}S}}{{{\rm{d}}A}} =  \frac{{\Delta {S_{\min }}}}{{\Delta {A_{\min }}}}
=  {\rm{ }}\frac{b}{{\lambda \ell _p^2}}{\left[ {1 + \frac{{\pi {\gamma ^2}}}{{4A}} + \mathcal{O}\left( {{\gamma ^3}} \right)} \right]^{ - 1}},
 \end{align}
Considering that $\gamma$ is a small parameter,  one can obtain the  GUP corrected entropy according to Eq.~(\ref{eq12+}),
\begin{align}
\label{eq13}
S = \frac{{A}}{{4\ell _p^2}}\left[ {1 - \frac{{\pi {\gamma ^2}}}{{4A}}\ln \left( {4A} \right) + \mathcal{O}\left( {{\gamma ^3}} \right)} \right],
\end{align}
where we f\/ix ${b \mathord{\left/ {\vphantom {b \lambda }} \right. \kern-\nulldelimiterspace} \lambda }$ by demanding matching the original entropy-area law in the limit $\gamma   \to 0$. Hence, one has ${b \mathord{\left/ {\vphantom {b \lambda }} \right. \kern-\nulldelimiterspace} \lambda } = { {1} \mathord{\left/ {\vphantom { {k_B} \pi}} \right. \kern-\nulldelimiterspace} 4}$  \cite{chy1}.  Moreover, there is a logarithmic term  $\ln \left( {4 A } \right)$ in the bracketed term of Eq.~(\ref{eq13}), which is coincident with the requirements of QG \cite{chb5,chb6,chb7,chf1,chf2,chb8,chg1,chg2}. According to the holographic principle and entropy-area law, which is averagely distributed in  $\mathcal{N}$-bits information, the number of bits can be expressed as follows:
\begin{align}
\label{eq14}
\mathcal{N}= 4S = \frac{A}{{\ell _p^2}}\left[ {1 - \frac{{\pi {\gamma ^2}}}{{4A}}\ln \left( {4A} \right) + \mathcal{O}\left( {{\gamma ^3}} \right)} \right].
\end{align}
Next, one can further denote the total energy of the gravitational system as  $E$, which is the average distribution in  $\mathcal{N}$ bits (please see the \cite{chg3} for details).  Consequently, each bit contains  ${{T} \mathord{\left/  {\vphantom {{{k_B}T} 2}} \right. \kern-\nulldelimiterspace} 2}$ energy, then, following the equipartition rule, the total energy takes the form
\begin{align}
\label{eq15}
E = {{\mathcal{N}T } \mathord{\left/ {\vphantom {{\mathcal{N}T} 2}} \right. \kern-\nulldelimiterspace} 2}.
\end{align}
On the other hand, Verlinde demonstrated that the entropic force is more fundamental than gravity. For a gravitational system, the entropic force can be expressed as follows \cite{cha28}:
\begin{align}
\label{eq16+}
F\Delta x = T\Delta S,
\end{align}
where  $F$ is the entropy force,   $T $ is the temperature,  $\Delta S$ is the change in entropy of the gravitational system, and  $\Delta x$ represents the displacement of the particle with the mass $m$  from the gravitational system, which satisf\/ies the relation $\Delta S = 2\pi m\Delta x$ \cite{chb2,chb3}. Now, substituting the expression of the change in entropy, Eq.~(\ref{eq13})-Eq.~(\ref{eq15}) into Eq.~(\ref{eq16+}), then considering $E = M{c^2}$, $A = 4\pi {r^2}$ and $\ell _p^2=G$, Newton's law of gravitation should be corrected as follows:
\begin{align}
\label{eq16}
{F_{{\rm{GUP}}}} =\frac{{GMm}}{{{r^2}}}\left[ {1 + \frac{{{\gamma ^2}}}{{16{r^2}}}\ln \left( {16\pi {r^2}} \right) + \mathcal{O}\left( {{\gamma ^3}} \right)} \right].
\end{align}
Clearly, in the limit  $\gamma =0$, Eq.~(\ref{eq16}) reduce to the original Newton's gravitation ${F_0} = {{GMm} \mathord{\left/  {\vphantom {{GMm} {{r^2}}}} \right. \kern-\nulldelimiterspace} {{r^2}}}$. Next, based on the modif\/ied Newton's law of gravitation, the corresponding GUP corrected gravitational potential is
\begin{align}
\label{eq17}
{V_{{\rm{GUP}}}} & = \int {\frac{{{F_{{\rm{GUP}}}}}}{m}dr}
 \nonumber \\
& =  - \frac{{GM}}{r}\left[ {1 + \frac{{2 + 3\ln \left( {16\pi {r^2}} \right)}}{{144{r^2}}}{\gamma ^2} + \mathcal{O}\left( {{\gamma ^3}} \right)} \right].
\end{align}
Applying Eq.~(\ref{eq17}) to a molecular cloud with radius  $R$, mass $M$  and the almost uniform density  $\rho_0$, the following expression for the modif\/ied potential energy is obtained
\begin{align}
\label{eq18}
E_p^{{\rm{GUP}}}&= \int_0^R {{V_{{\rm{GUP}}}}\left( r \right)dM}
 \nonumber \\
& =  - \frac{{3G{M^2}}}{{5R}}\left[ {1 + \frac{{5\ln \left( {16\pi {R^2}} \right)}}{{144{R^2}}}{\gamma ^2} + \mathcal{O}\left( {{\gamma ^3}} \right)} \right].
\end{align}
Obviously, the above equation is related to the mass  $M$, radius  $R$, and GUP parameter  $\gamma$. In addition, one can see that Eq.~(\ref{eq18}) lacks the f\/irst-order correction term since the CH model~(\ref{eq2}) becomes the KMM model for small $\gamma$. This indicates that the properties of the gravitational system are af\/fected by the higher-order term of $\gamma$. In the limit $\gamma =0$, the result  agrees with the original potential energy~(\ref{eq4}).

\subsection{GUP corrected canonical energy of ideal gases}
By virtue of Eq.~(\ref{eq2}) and the viewpoint in Refs.~\cite{chb3+,chc11},  the partition function of a dark cloud, which is approximated as the ideal gas containing $N$  noninteracting particles with mass $m$  at temperature $T$, can be expressed as follows:
\begin{align}
\label{eq19}
Z = \frac{{{\mathcal{Z}^N}}}{{N!}},
\end{align}
where
\begin{align}
\label{eq19}
\mathcal{Z}= \frac{{4\pi V}}{{N!}}\int_0^\infty  {\frac{{{p^2}}}{{{h^3}}}{{\left( {1 - 3\gamma \left| p \right|} \right)}^3}\exp \left( { - \frac{{{p^2}}}{{2\mu T}}} \right)dp}.
\end{align}
Using the spherical coordinate and Gaussian integral, the GUP corrected partition function for the dark cloud reads
\begin{align}
\label{eq20}
{Z_{{\rm{GUP}}}} = \frac{{{{\left( {4\pi V} \right)}^N}{{\left( {2{k_B} \mu T} \right)}^{3N/2}}}}{{N!{h^{3N}}}}{\Theta ^N}\left( \gamma  \right),
\end{align}
where $\Theta \left( \gamma  \right) = \frac{{\sqrt \pi  }}{4}\left[ {1 - 6\sqrt {\frac{{2 \mu T}}{\pi }} \gamma  + 9 \mu T{\gamma ^2} + \mathcal{O}\left( {{\gamma ^3}} \right)} \right]$. The GUP corrected canonical energy is presented as
\begin{align}
\label{eq22}
U_{{\rm{GUP}}}  = & {T^2}{\left( {\frac{{\partial \ln {Z_{{\rm{GUP}}}}}}{{\partial T}}} \right)_{N,V}}
 \nonumber \\
= &  {U_0}\left[ {1 - 2\sqrt {\frac{2}{\pi }} {\cal A}\gamma  - \frac{6}{\pi }\left( {4 - \pi } \right){{\cal A}^2}{\gamma ^2} + {\cal O}\left( {{\gamma ^3}} \right)}\right],
\end{align}
with $\mathcal{A} =\sqrt {\mu T} $ and the original canonical energy ${U}_0$ is given in Eq.~(\ref{eq7}). Note that, the modif\/ied canonical energy is dif\/ferent from the original case, it does only not relate to the original case  $U_0$, but also to parameter $\gamma $, the mass of the noninteracting particle $\mu$  and its temperature  $T$.

\subsection{GUP corrected Jeans mass}
 With the above results in place, one can derive the modif\/ied Jeans mass and  further investigate how the GUP af\/fects the  Jeans mass.  Using Eq.~(\ref{eq18}) and Eq.~(\ref{eq22}) and recalling Eq.~(\ref{eq3}), one yields
\begin{align}
\label{eq23}
NT&  \left[{1 - 2\sqrt {\frac{2}{\pi }} {\cal A}\gamma  - \frac{6}{\pi }\left( {4 - \pi } \right){{\cal A}^2}{\gamma ^2}} { + {\cal O}\left( {{\gamma ^3}} \right)}\right] <
 \nonumber \\
&\frac{{G{M^2}}}{{5R}} \left[ {1 + \frac{{5\ln \left( {16\pi {R^2}} \right)}}{{144{R^2}}}{\gamma ^2} + \mathcal{O}\left( {{\gamma ^3}} \right)} \right].
\end{align}
Considering the radius $R = {\left( {{{3M} \mathord{\left/ {\vphantom {{3M} {4\pi {\rho _0}}}} \right. \kern-\nulldelimiterspace} {4\pi {\rho _0}}}} \right)^{\frac{1}{3}}}$, Eq.~(\ref{eq23}) can be rewritten as follows:
\begin{align}
\label{eq24}
\frac{{5T\left[{1 - 2\sqrt {\frac{2}{\pi }} {\cal A}\gamma  - \frac{6}{\pi }\left( {4 - \pi } \right){{\cal A}^2}{\gamma ^2}} { + {\cal O}\left( {{\gamma ^3}} \right)} \right]}}{{G \mu {{\left( {{{4\pi {\rho _0}} \mathord{\left/  {\vphantom {{4\pi {\rho _0}} 3}} \right. \kern-\nulldelimiterspace} 3}} \right)}^{\frac{1}{3}}}}} < {M^{\frac{2}{3}}}\chi  ,
\end{align}
where $ \chi= {1 + \frac{{5{\gamma ^2}}}{{144{{\left( {{{3M} \mathord{\left/ {\vphantom {{3M} {4\pi {\rho _0}}}} \right. \kern-\nulldelimiterspace} {4\pi {\rho _0}}}} \right)}^2}}}\ln \left[ {16\pi {{\left( {{{3M} \mathord{\left/ {\vphantom {{3M} {4\pi {\rho _0}}}} \right. \kern-\nulldelimiterspace} {4\pi {\rho _0}}}} \right)}^2}} \right] + \mathcal{O}\left( {{\gamma ^3}} \right)}$. On the one hand, if  $\gamma =0$, the upper bound of mass goes to the Jeans mass  $M_{\rm{0}}^J = {\left( {{{5 T} \mathord{\left/ {\vphantom {{5 T} {G \mu}}} \right.\kern-\nulldelimiterspace} {G \mu}}} \right)^{\frac{3}{2}}}{\left( {{3 \mathord{\left/ {\vphantom {3 {4\pi {\rho _0}}}} \right. \kern-\nulldelimiterspace} {4\pi {\rho _0}}}} \right)^{\frac{1}{2}}}$. On the other hand, because $M \gg 1$ and $\gamma $  is f\/inite, the GUP corrected Jeans mass is obtained by saturating Eq.~(\ref{eq24}), with the result
\begin{align}
\label{eq25}
&M_{{\rm{GUP}}}^J  = M
 \nonumber \\
 & > M_{\rm{0}}^J{\left[ {1 - \sqrt {\frac{{8\mu T}}{\pi }} \gamma  - \left( {4 - \pi } \right)\frac{{6\mu T}}{\pi }{\gamma ^2} + \mathcal{O}\left( {{\gamma ^3}} \right)} \right]^{\frac{3}{2}}}.
\end{align}
One may see that the GUP corrected Jeans mass  $M_{{\rm{GUP}}}^J$ is related to the original Jeans mass  $M_0^J$, the mass of ideal gas $\mu$, the temperature of gravity system $T$, and the GUP parameter  $\gamma $. Furthermore, it should be noted that the mass of ideal gas and the temperature of the gravity system must be real numbers greater than zero. When considering $\gamma>0$, $M_{{\rm{GUP}}}^J$ is positive when $1 > \sqrt {{{8\mu T} \mathord{\left/ {\vphantom {{8\mu T} \pi }} \right. \kern-\nulldelimiterspace} \pi }} \gamma  - \left( {4 - \pi } \right){{6\mu T{\gamma ^2}} \mathord{\left/ {\vphantom {{6\mu T{\gamma ^2}} \pi }} \right. \kern-\nulldelimiterspace} \pi } + {\cal O}\left( {{\gamma ^3}} \right)$ is lower than $M_{\rm{0}}^J$. To date, the astronomical observations have shown that some Bok globules' masses are less than their corresponding Jeans masses. The GUP corrections to the Jeans mass can be a candidate to explain those observational facts. Our results show that the ef\/fect of GUP is able to reduce the Jeans mass, which is conducive to the generation of stars in small-mass Bok globules.

\section{Constraints for GUP parameter $\gamma _0$}
\label{sec4}
The GUP parameters  are always assumed to be of order unity so that the modif\/ied results are negligible unless energy approaches the Planck scale. However, if the assumptions regarding the GUP parameters are not considered, the bound of GUP parameters can be obtained by previous experimental and observational data. In previous works, people mainly focused on the constraints on the GUP parameters of the KMM model and the ADD model \cite{chc1,chc2,chc3,chc4,chc5,chc5+,chy5,chc10,chc7,cha30,chc6,chc8,cha31,chc9,chc10+,chy10,chy11}. Now, armed with the previous results, we estimate the parameter of the CH model based on the data of Bok globules. First, for simplicity, by keeping the leading order term of  $\gamma$, inequality~(\ref{eq25}) can be written as follows
\begin{align}
\label{eq26}
M_{{\rm{GUP}}}^J > M_{\rm{0}}^J{\left( {1 - \sqrt {\frac{{8\mu T}}{\pi }} \frac{\gamma _0}{{{M_p}c}}} \right)^{\frac{3}{2}}}.
\end{align}
By solving the above equation, one yields
\begin{align}
\label{eq26+}
\gamma _0  < \frac{{c{M_p}}}{2}\sqrt {\frac{\pi }{{2\mu T}}} \left[ {1 - {{\left( {\frac{{M_{{\rm{GUP}}}^J}}{{M_{\rm{0}}^J}}} \right)}^{2/3}}} \right].
\end{align}
Now, based on the data of Bok globules in Ref.~\cite{cha17}, the upper bounds on $\gamma _0$ are shown in Table.~\ref{tab1}.
\begin{table*}[htb]
\caption {\label{tab1} The ID, temperature $T$, mass $M_{{\rm{GUP}}}^J$ and  Jeans mass ${M_0^J}$ of Bok globules,  $\gamma _0$ and $\gamma _0^2$ are the f\/irst power and second power of the upper bound on the GUP parameter of the CH model, $\beta _0^{\rm {KMM}}$ is the upper bound on the GUP parameter of the KMM model.}

\centering
\begin{tabular}{c c c c c  c  c}
 \hline
ID         &    $ T ({\rm{K}}) $\cite{cha17}    &   $M_{{\rm{GUP}}}^J ({{{\rm{M}}_ \odot }})$  \cite{cha17}        &       ${M_0^J} ({{{\rm{M}}_ \odot }})$ \cite{cha17}     &  $\gamma _0$   &   ${\gamma _0^2}$  & $\beta _0^{\rm {KMM}}$ \cite{cha19}  \\
 \hline
CB 87                          &    11.4            &    2.73                       &         9.6             &   $5.31 \times {10^{12}}$        &   $2.82 \times {10^{24}}$     & $6.33 \times {10^{25}}$\\
CB 110                       &    21.8            &    7.21                        &         8.5            &   $7.03 \times {10^{11}}$        &   $5.00 \times {10^{23}}$     & $6.11\times {10^{24}}$\\
CB 131                       &    25.1            &    7.83                       &         8.1             &   $1.41 \times {10^{11}}$        &   $1.99 \times {10^{22}}$     & $1.02 \times {10^{24}}$\\
CB 161                       &    12.5            &    2.79                       &         5.4             &   $3.18 \times {10^{12}}$        &   $1.01 \times {10^{25}}$     &$3.62 \times {10^{26}}$\\
CB 188                       &    19.0            &    7.19                       &         7.7            &   $3.24 \times {10^{11}}$        &   $1.05 \times {10^{23}}$     &$3.05 \times {10^{24}}$\\
FeSt 1-457                &    10.9            &    1.12                       &         1.4            &   $1.32 \times {10^{12}}$        &   $1.75 \times {10^{24}}$     &$1.60\times {10^{25}}$\\
Lynds 495                &    12.6            &    2.95                        &         6.6            &   $1.70 \times {10^{12}}$        &   $1.37 \times {10^{25}}$     &$4.20 \times {10^{25}}$\\
Lynds 498                &    11.0            &    1.42                        &         5.7            &   $5.76\times {10^{12}}$         &   $3.31\times {10^{25}}$      &$6.99 \times {10^{25}}$\\
 \hline
 \end{tabular}
\end{table*}

In Table.~\ref{tab1}, we set $M_p=2.18\times {10^{-8}}{\rm kg}$, $c = 2.99 \times {10^8}{\rm{m}}\cdot{{\rm{s}}^{ - 1}}$, and ${{\rm{M}}_ \odot }$ denotes the Sun mass. The f\/inal results are determined by the mass of noninteracting particles  $\mu$. To obtain the exact value of the upper bounds on  $\gamma _0$, we further assume that the noninteracting particles are composed of hydrogen atoms, which are the most abundant element in the universe with the mass  ${\mu _{\rm{H}}} = 1.67 \times {10^{- 26}}{\rm kg}$. In this case, one can see that the range of upper bounds of $\gamma _0$ is ${10^{11}} \sim {10^{12}}$. Furthermore, the GUP parameter in the CH model has a special relationship with that of the KMM model, which can be inferred from Eq.~(\ref{eq2}), as ${\gamma ^2} \sim \beta$ (or $\gamma _0^2 \sim {\beta _0}$). Hence, it is easy to f\/ind that the upper bound of the GUP parameter in the CH model $\gamma _0^2$ is 1-2 orders of magnitude more stringent than those in the KMM model $\beta _0^{\rm {KMM}} < {{\left[ {1 - {{\left( {{{M_{GUP}^J} \mathord{\left/ {\vphantom {{M_{GUP}^J} {M_0^J}}} \right. \kern-\nulldelimiterspace} {M_0^J}}} \right)}^{{2 \mathord{\left/ {\vphantom {2 3}} \right.  \kern-\nulldelimiterspace} 3}}}} \right]M_p^2 c^2} \mathord{\left/ {\vphantom {{\left[ {1 - {{\left( {{{M_{GUP}^J} \mathord{\left/ {\vphantom {{M_{GUP}^J} {M_0^J}}} \right. \kern-\nulldelimiterspace} {M_0^J}}} \right)}^{{2 \mathord{\left/ {\vphantom {2 3}} \right. \kern-\nulldelimiterspace} 3}}}} \right]M_p^2 c^2} {2\mu T}}} \right. \kern-\nulldelimiterspace} {2\mu T}}$ from Ref.~\cite{cha19}.

\section{Discussion}
\label{sec5}
In this paper, by incorporating a new higher-order GUP with virial equilibrium and Verlinde's entropy force theory, we investigated the modif\/ied canonical energy $U_{{\rm{GUP}}} $ and the modified gravitational potential energy $E_p^{{\rm{GUP}}}$ of a molecular cloud. Subsequently, according to those modif\/ications, the GUP corrected Jeans mass $M_{{\rm{GUP}}}^J $ is obtained. It is found that the GUP can ef\/fectively increase the gravitational potential, and reduce the canonical energy  and the Jeans mass. This leads to the collapse of Bok globules with masses less than the standard value  $M_{\rm{0}}^J$, which is consistent with astronomical observations. Finally, using the dif\/ferent data of Bok globules, we constrain the upper bounds of the GUP parameter  $\gamma _0$, whose range turn out to be ${10^{11}} \sim {10^{12}}$.


\begin{thebibliography}{99}
\bibitem{chb1}
P. Chen, Y. C. Ong, D.-H. Yeom, Phys. Rep. \textbf{603},  1 (2015). \href{https://arxiv.org/abs/1412.8366} {arXiv:1412.8366}

\bibitem{cha1}
M. Maggiore, Phys. Lett. B \textbf{319},  83 (1993). \href{https://arxiv.org/abs/hep-th/9309034} {arXiv:hep-th/9309034}

\bibitem{cha2}
G. Amelino-Camelia, Int. J. Mod. Phys. D \textbf{11}, 35 (2002).  \href{https://arxiv.org/abs/gr-qc/0012051} {arXiv:gr-qc/0012051}

\bibitem{cha3}
 F. Scardigli, Phys. Lett. B \textbf{452},   39 (1999). \href{https://arxiv.org/abs/hep-th/9904025} {arXiv:hep-th/9904025}

\bibitem{cha4}
A. Kempf, G. Mangano, R. B. Mann, Phys. Rev. D \textbf{52}, 1108 (1995). \href{https://arxiv.org/abs/hep-th/9412167} {arXiv:hep-th/9412167}

\bibitem{cha5}
A. F. Ali, S. Das, E. C. Vagenas, Phys. Lett. B \textbf{678},  497 (2009). \href{http://dx.doi.org/ 10.1016/j.physletb.2009.06.061} {DOI: 10.1016/j.physletb.2009.06.061}

\bibitem{cha5+}
G. Lambiase, F. Scardigli, Phys. Rev. D \textbf{97},  075003 (2018). \href{https://arxiv.org/abs/1709.00637} {arXiv:1709.00637}

\bibitem{cha5++}
R. Casadio, F. Scardigli, Phys. Lett. B \textbf{807},  135558 (2020). \href{https://arxiv.org/abs/2004.04076} {arXiv:2004.04076}

\bibitem{cha6}
Z. W. Feng, H. L. Li, X. Z. Tao, S. Z. Yang, Eur. Phys. J. C  \textbf{76},  212 (2016). \href{https://arxiv.org/abs/1604.04702} {arXiv:1604.04702}

\bibitem{cha7}
I. Sakalli, A. \"{O}vg\"{u}n, 	Europhys. Lett. \textbf{118},  60006 (2017). \href{https://arxiv.org/abs/1702.04636} {arXiv:1702.04636}

\bibitem{cha8}
A. \"{O}vg\"{u}n, K. Jusuf\/i, Eur. Phys. J. Plus   \textbf{132},  298 (2017).  \href{https://arxiv.org/abs/1703.08073} {arXiv:1703.08073}

\bibitem{cha8+}
F. Scardigli, M. Blasone, G. Luciano, R. Casadio, Eur. Phys. J. C \textbf{78}, 728 (2018). \href{http://dx.doi.org/10.1140/epjc/s10052-018-6209-y} {DOI: 10.1140/epjc/s10052-018-6209-y}

\bibitem{cha9}
E. C. Vagenas, A. F. Ali, H. Alshal, Eur. Phys. J. C \textbf{79},   276 (2019). \href{https://arxiv.org/abs/1811.06614} {arXiv:1811.06614}

\bibitem{cha10}
K. Jusuf\/i, P. Channuie, M. Jamil, 	Eur. Phys. J. C \textbf{80},  127 (2020). \href{https://arxiv.org/abs/2002.01341} {arXiv:2002.01341}

\bibitem{cha10+}
M. Blasone, G. Lambiase, G. Luciano, L. Petruzziello, F. Scardigli, Int. J. Mod. Phys. D \textbf{29}, 2050011 (2020). \href{http://dx.doi.org/10.1142/S021827182050011X} {DOI: 10.1142/S021827182050011X}

\bibitem{cha11}
P. Pedram, Phys. Lett. B \textbf{714},  317 (2012). \href{https://arxiv.org/abs/1110.2999} {arXiv:1110.2999}

\bibitem{cha12}
P. Pedram, Phys. Lett. B \textbf{718}, 638 (2012). \href{http://dx.doi.org/10.1016/j.physletb.2012.10.059} {DOI: 10.1016/j.physletb.2012.10.059}

\bibitem{cha13}
W. S. Chung, H. Hassanabadi, Eur. Phys. J. C  \textbf{79},  213 (2019). \href{https://doi.org/10.1140/epjc/s10052-019-6718-3} {DOI: 10.1140/epjc/s10052-019-6718-3}

\bibitem{cha14}
H. Hassanabadi, E. Maghsoodi, W. S. Chung, Eur. Phys. J. C  \textbf{79}, 358 (2019). \href{https://doi.org/10.1140/epjc/s10052-019-6871-8} {DOI: 10.1140/epjc/s10052-019-6871-8}

\bibitem{cha15}
H. Shababi, W. S. Chung, Mod. Phys. Lett. A \textbf{33}, 1850068 (2018). \href{https://doi.org/10.1142/S0217732320500182} {DOI: 10.1142/S0217732320500182}

\bibitem{cha15+}
H. Shababi, W. S. Chung, Phys. Lett. B \textbf{770},  445 (2017). \href{https://doi.org/10.1016/j.physletb.2017.05.015} {DOI:  10.1016/j.physletb.2017.05.015}

\bibitem{cha16}
W. S. Chung,  H. Hassanabadi,  Eur. Phys. J. C \textbf{79}, 213  (2019). \href{https://doi.org/10.1140/epjc/s10052-019-6718-3} {DOI:  10.1140/epjc/s10052-019-6718-3}

\bibitem{cha17}
J. Vainio, I. Vilja,  Gen. Relativ. Gravit. \textbf{48},  129 (2016). \href{https://arxiv.org/abs/1512.04220} {arXiv:1512.04220}

\bibitem{cha18}
R. Kandori,  Y. Nakajima, M. Tamura, K. Tatematsu, Y. Aikawa, T. Naoi, K. Sugitani, H. Nakaya, T. Nagayama, T. Nagata, M. Kurita, D. Kato, C. Nagashima, S. Sato, Astron. J. \textbf{130}, 2166 (2005). \href{https://arxiv.org/abs/astro-ph/0506205}  {arXiv: astro-ph/0506205}

\bibitem{cha19}
H. Moradpour, A. H. Ziaie, S. Ghaf\/fari, F. Feleppa, Mon. Not.  R  Astron. Soc. \textbf{488},  L69 (2019). \href{https://arxiv.org/abs/1907.12940}  {arXiv:1907.12940}

\bibitem{che1}
L. N. Chang, D. Minic, N. Okamura, T. Takeuchi, Phys. Rev. D \textbf{65},  125028 (2002). \href{https://arxiv.org/abs/hep-th/0201017}  {arXiv:hep-th/0201017}

\bibitem{cha20}
P. Wang, H. Yang, X. Zhang, J. High Energy Phys. \textbf{1008},  043 (2010). \href{https://arxiv.org/abs/1006.5362}  {arXiv:1006.5362}

\bibitem{cha21}
W. S. Chung, H. Hassanabadi, Int. J.  Mod. Phys. A \textbf{34},   1950041 (2019).  \href{https://doi.org/10.1142/S0217751X19500416} {DOI: 10.1142/S0217751X19500416}

\bibitem{cha22}
Z.-W. Feng, S.-Z. Yang, H.-L. Li, X.-T. Zu, Adv. High Energy Phys. \textbf{2016},   2341879 (2016).  \href{https://arxiv.org/abs/1607.04114}  {arXiv:1607.04114}

\bibitem{cha23}
R. J. Adler, P. Chen, D. I. Santiago, Gen. Relativ. Gravit. \textbf{33},  2101 (2001).   \href{http://dx.doi.org/10.1023/A:1015281430411} {DOI: 10.1023/A:101528143041}

\bibitem{chd1}
M. Cavaglia, S. Das, R. Maartens, Class. Quantum Grav. \textbf{20},  L205 (2003). \href{https://arxiv.org/abs/hep-ph/0305223} {arXiv:hep-ph/0305223}

\bibitem{chd2}
A. J. M.Medved, E. C. Vagenas, Phys. Rev. D \textbf{70}, 12402 (2004).  \href{https://arxiv.org/abs/hep-th/0411022} {arXiv:hep-th/0411022}

\bibitem{chd3}
G. Amelino-Camelia, M. Arzano, A. Procaccini, Phys. Rev. D \textbf{70},  107501 (2004). \href{https://arxiv.org/abs/gr-qc/0405084} {arXiv:gr-qc/0405084}

\bibitem{chd3+}
I. H. Belfaqih, H. Maulana, A. Sulaksono, Int. J Mod. Phys. D  \href{http://dx.doi.org/10.1142/S0218271821500644} {DOI: 10.1142/S0218271821500644}

\bibitem{cha28}
E. Verlinde, J. High Energy Phys. \textbf{2011},  29 (2011).  \href{https://arxiv.org/abs/1001.0785}{arXiv:1001.0785}

\bibitem{chy1}
A. Awad, A. F. Ali, Cent. Eur. J Phys. \textbf{12},  245 (2014).  \href{https://arxiv.org/abs/1403.5319}{arXiv:1403.5319}

\bibitem{cha24}
J. D. Bekenstein, Phys. Rev. D \textbf{7}, 2333 (1973). \href{http://dx.doi.org/10.1103/PhysRevD.7.2333} {DOI: 10.1103/PhysRevD.7.2333}

\bibitem{chy2}
A. Awad,  A. F. Ali, J. High Energy Phys. \textbf{2014},  93 (2014).  \href{https://arxiv.org/abs/1404.7825}{arXiv:1404.7825}

\bibitem{cha25}
G. Amelino-Camelia, M. Arzano, A. Procaccini, Phys. Rev. D \textbf{70},  107501 (2004).  \href{https://arxiv.org/abs/gr-qc/0405084} {arXiv:gr-qc/0405084}

\bibitem{cha26}
B. Majumder, Phys. Lett. B \textbf{703}, 402 (2011). \href{https://arxiv.org/abs/1106.0715} {arXiv:1106.0715}

\bibitem{chy3}
A. J. M. Medved, E. C. Vagenas, Phys. Rev. D \textbf{70}, 124021 (2004). \href{https://arxiv.org/abs/hep-th/0411022} {arXiv:hep-th/0411022}

\bibitem{chy4}
B. Majumder, Phys. Lett. B \textbf{703}, 402 (2011). \href{https://arxiv.org/abs/1106.0715} {arXiv:1106.0715}

\bibitem{chy4+}
R. A. El-Nabulsi, Quantum Stud.: Math. Found. \textbf{6}, 235 (2019). \href{https://doi.org/10.1007/s40509-019-00181-x} {DOI: 10.1007/s40509-019-00181-x}

\bibitem{cha27}
C. Adami, \href{https://arxiv.org/abs/quant-ph/0405005} {arXiv:quant-ph/0405005}

\bibitem{chb5}
P. Bargue\={n}o, E. C. Vagenas, Phys. Lett. B, \textbf{742}, 15 (2015). \href{https://arxiv.org/abs/1501.03256}  {arXiv:1501.03256}

\bibitem{chb6}
N. M.-Dur\'{a}n, A. F. Vargas, P. Hoyos-Restrepo, P. Bargue\={n}o, Eur. Phys. J. C \textbf{76},   559 (2016). \href{https://arxiv.org/abs/1606.06635}{arXiv:1606.06635}

\bibitem{chb7}
Z.-W. Feng, S.-Z. Yang, Phys. Lett. B  \textbf{772},  737 (2017). \href{https://arxiv.org/abs/1501.03256}{arXiv:1501.03256}

\bibitem{chf1}
 Z. F. Gao, X-D. Li, N. Wang, J. P. Yuan, P. Wang, Q. H. Peng, Y. J. Du, Mon. Not. R. Astron. Soc. \textbf{456}, 55 (2016).  \href{http://dx.doi.org/10.1093/mnras/stv2465} {DOI: 10.1093/mnras/stv2465}

 \bibitem{chf2}
Z. F. Gao, D. L. Song, X. D. Li, H. Shan, N. Wang, Astron. Nachr.  \textbf{340}, 241 (2019). \href{http://dx.doi.org/10.1002/asna.201913599} {DOI: 10.1002/asna.201913599}

\bibitem{chb8}
Z.-Y. Fu, H.-L. Li, Y. Li, D.-W. Song, Eur. Phys. J. Plus \textbf{135},  125 (2020).  \href{http://dx.doi.org/10.1140/epjp/s13360-020-00190-5} {DOI: 10.1140/epjp/s13360-020-00190-5}

\bibitem{chg1}
F. Scardigli, Symmetry \textbf{12}, 1519 (2020). \href{http://dx.doi.org/10.1002/asna.201913599} {DOI: 10.3390/sym12091519}

\bibitem{chg2}
A. Iorio, G. Lambiase, P. Pais, F. Scardigli,  Phys. Rev. D \textbf{101}, 105002 (2020).  \href{https://arxiv.org/abs/1910.09019} {arXiv:1910.09019}

\bibitem{chg3}
F. Scardigli,  Class. Quantum Grav. \textbf{14}, 1781 (1997). \href{https://arxiv.org/abs/gr-qc/9706030} {arXiv:gr-qc/9706030}

\bibitem{chb2}
I. Sakalli, 	Int. J. Theor. Phys. \textbf{50}, 2426  (2011).  \href{https://arxiv.org/abs/1103.1728} {arXiv:1103.1728}

\bibitem{chb3}
A. Sheykhi, H. Moradpour, N. Riazi, Gen. Relativ. Gravit. \textbf{45}, 1033 (2013). \href{https://arxiv.org/abs/1109.3631}{arXiv:1109.3631}

\bibitem{chb3+}
B. Hamil, B. C. L\"{u}tf\"{u}o\u{g}lu, \href{https://arxiv.org/abs/2009.13838}{arXiv:2009.13838}

\bibitem{chc11}
 R. A. El-Nabulsi, Eur. Phys. J. Plus \textbf{135}, 34 (2020). \href{http://dx.doi.org/10.1140/epjp/s13360-019-00051-w} {DOI: 10.1140/epjp/s13360-019-00051-w}

\bibitem{chc1}
S. Das, E. C. Vagenas, Phys. Rev. Lett. \textbf{101},   221301 (2008).   \href{https://doi.org/10.1103/PhysRevLett.101.221301} {DOI: 10.1103/PhysRevLett.101.221301}

\bibitem{chc2}
F. Marin, F. Marino, M. Bonaldi, M. Cerdonio, L. Conti, P. Falferi, R. Mezzena, A. Ortolan, G. A. Prodi, L. Taf\/farello, G. Vedovato, A. Vinante, J. P. Zendri, Nat. Phys. \textbf{9},  71 (2013). \href{http://dx.doi.org/10.1038/nphys2503} {DOI: 10.1038/nphys2503}

\bibitem{chc3}
S. Ghosh,   Class. Quantum Grav. \textbf{31},  025025 (2014). \href{https://arxiv.org/abs/1303.1256}{arXiv:1303.1256}

\bibitem{chc4}
F.  Scardigli,  R.  Casadio, Eur. Phys. J. C \textbf{75},  425 (2015).  \href{https://arxiv.org/abs/1407.0113}{arXiv:1407.0113}

\bibitem{chc5}
D. Gao, M. Zhan, Phys. Rev. A \textbf{94},   013607 (2016). \href{https://arxiv.org/abs/1607.04353}{arXiv:1607.04353}

\bibitem{chc5+}
F. Scardigli, G. Lambiase, E. C. Vagenas, Phys. Lett. B  \textbf{767},  242 (2017). \href{https://arxiv.org/abs/1709.00637}{arXiv:1709.00637}

\bibitem{chy5}
F. Scardigli, J. Phys. Conf. Ser. \textbf{1275}, 012004 (2019). \href{https://arxiv.org/abs/1905.00287}{arXiv:1905.00287}

\bibitem{chc10}
 A. F. Ali, S. Das, E. C. Vagenas, Phys. Rev. D \textbf{84},  044013 (2011).  \href{https://arxiv.org/abs/1107.3164}{arXiv:1107.3164}

 \bibitem{chc7}
 P. Bushev,  J. Bourhill,  M. Goryachev,  N. Kukharchyk, E. Ivanov, S. Galliou, M. Tobar, S. Danilishin, Phys. Rev. D \textbf{100},  066020 (2019).  \href{https://arxiv.org/abs/1903.03346}{arXiv: 1903.03346}

\bibitem{cha30}
Z.-W. Feng, S.-Z. Yang, H.-L. Li, X.-T. Zu, Phys. Lett. B  \textbf{768},  81 (2017). \href{https://arxiv.org/abs/1610.08549}{arXiv:1610.08549}

\bibitem{chc6}
S. Kouwn, Phys. Dark Universe  \textbf{21},  76 (2018). \href{https://arxiv.org/abs/1805.07278}{arXiv:1805.07278}

\bibitem{chc8}
S.  Giardino,  V.  Salzano,  \href{https://arxiv.org/abs/2006.01580}{arXiv:2006.01580}

\bibitem{cha31}
J. C. S. Neves, Eur. Phys. J. C \textbf{80},   343 (2020). \href{https://arxiv.org/abs/1906.11735}{arXiv:1906.11735}

\bibitem{chc9}
S.  Das  and  R.  Mann, Phys. Lett. B  \textbf{704},  596 (2011). \href{https://arxiv.org/abs/1109.3258}{arXiv:1109.3258}

 \bibitem{chc10+}
 S. Bhattacharyya, S. Gangopadhyay, A. Saha, Class. Quantum Grav. \textbf{37},  195006  (2020).  \href{http://dx.doi.org/10.1088/1361-6382/abac45} {DOI: 10.1088/1361-6382/abac45}

 \bibitem{chy10}
\"{O}. \"{O}kc\"{u},  E. Aydiner, Nucl. Phys. B \textbf{964}, 115324 (2021). \href{https://arxiv.org/abs/2101.09524}{arXiv:2101.09524}

\bibitem{chy11}
A. Das,  S. Das,  N. R. Mansour,  E. C. Vagenas, Phys. Lett. B \textbf{819}, 136429 (2021).   \href{https://arxiv.org/abs/2101.03746}{arXiv:2101.03746}
\end{thebibliography}
\end{document}